\documentclass[]{interact}

\usepackage{epstopdf}
\usepackage[caption=false]{subfig}

\usepackage[natbibapa,nodoi]{apacite}
\theoremstyle{plain}

\theoremstyle{definition}

\theoremstyle{remark}

\PassOptionsToPackage{hyphens}{url}\usepackage{hyperref} 

\usepackage{comment}

\usepackage{dcolumn}

\usepackage{verbatim}

\begin{document}

\articletype{ARTICLE}

\title{Unleashing Data Journalism's Potential: COVID-19 as Catalyst for Newsroom Transformation}


\author{
\name{Benedict Witzenberger\thanks{CONTACT Benedict Witzenberger. Email: benedict.witzenberger@tum.de} and J\"urgen Pfeffer}
\affil{Technical University of Munich, Munich, Germany}
}

\maketitle

\begin{abstract}
In the context of journalism, the COVID-19 pandemic brought unprecedented challenges, necessitating rapid adaptations in newsrooms. Data journalism emerged as a pivotal approach for effectively conveying complex information to the public. Here, we show the profound impact of COVID-19 on data journalism, revealing a surge in data-driven publications and heightened collaboration between data and science journalists. 
Employing a quantitative methodology, including negative binomial regression and Relational hyperevent models (RHEM), on byline data of articles co-authored by data journalists, we comprehensively analyze data journalism outputs, authorship trends, and collaboration networks to address five key research questions.

The findings reveal a significant increase in data journalistic pieces during and after the pandemic, in particular with a rise in publications within scientific departments. Collaborative efforts among data and science journalists intensified, evident through increased authorship and co-authorship trends. Prior common authorship experiences somewhat influenced the likelihood of future co-authorships, underscoring the importance of building collaborative communities of practice.

These quantitative insights provide an understanding of the transformational role of data journalism during COVID-19, contributing to the growing body of literature in computational communication science and journalism practice.
\end{abstract}



\section{Introduction}

The COVID-19 pandemic has brought unprecedented challenges to the field of journalism, compelling newsrooms to adapt swiftly to the rapidly evolving information landscape \citep{Quandt2021,Mellado2021,Hanusch2022}. In the wake of this crisis, data journalism has emerged as a powerful and vital approach to communicating complex information to the public \citep{Pentzold2019,DanzonChambaud2021,GarciaAviles2022}. This article explores the profound impact of COVID-19 on data journalism, focusing on the observable surge in data-driven publications and the heightened collaboration between data and science journalists. 

Prior qualitative research has provided indications that data journalism gained deeper inclusion in newsrooms during the pandemic \citep{Wu2021,Bisiani2023}. These preliminary findings suggest that the COVID-19 crisis may have served as a catalyst for news organizations to recognize the value and importance of data-driven reporting in effectively communicating critical information to the public. The qualitative insights have highlighted how newsrooms embraced data journalism as a means to make sense of complex data related to the pandemic, enabling them to provide audiences with accurate, visually engaging, and accessible information \citep{Pentzold2021}.

To strengthen these findings, we will adopt a quantitative approach. This study seeks to complement prior qualitative research by systematically analyzing data journalism outputs, authorship patterns, and collaboration networks on a larger scale, using byline data — the short text snippet identifying the author of a text to allow attribution to the individual responsible for the piece — from articles that data journalists co-authored.
We will regard two time periods: pre-COVID-19 — the time before the pandemic appeared, and post-COVID-19, which does not imply the eradication of the virus but the time period in the aftermath of the global occurrence of COVID-19.
Results will be validated using common statistical inferences and modeling tools, namely $\chi^2$ tests and negative binomial regression models. To be able to also model author cooperations based on historical data, we will call on Relational hyperevent models (RHEM) that help explain network evolution in relational event history data — where an article is regarded as an event in this study.

We will ground this work on prior qualitative research on the influence of COVID-19 on data journalism and aim to embed our findings within Communities of Practice (CoP) that are often used to explain knowledge sharing in group contexts. Communities of Practice can be described as groups of people who share a common interest and collaborate to learn from one another, develop their skills, and share knowledge and expertise within that specific domain. We argue that the collaboration between data journalists and their colleagues in the newsroom might be a form of common learning and sharing of knowledge during the pandemic.

\subsection{COVID-19 and Data Journalism}

Firstly, we want to provide some background on data journalism and its development during COVID-19 to show the gap in research that this paper aims to fill.

Data journalism is a young profession. It can be dated back to the first decade of the 21st century \citep{Bravo2020}. However, its roots can be traced back to social-science methods proposed for precision journalism \citep{Meyer1973} and computer-assisted reporting (CAR), with which it shares some connection to investigative reporting \citep{Coddington2015}. Since the early 2010s, data-driven storytelling has been on the rise around the world \citep{Segel2010,Rogers2011,Hermida2019}, mostly in large, well-staffed news organizations \citep{Beiler2020,Haim2022}, but also occurs in local settings, with lower staffing \citep{Stalph2022}. Three factors have been found to be central in shaping data journalism \citep{Appelgren2019}: journalistic cultures define to what extend watchdog-transparency is regarded as a central value for strengthening the governing political system — democracies regard transparency to its processes and actors as central, while other autocratic systems may not \citep{Hanitzsch2019,Lewis2019}. This is a further factor regarding the political systems that data journalists operate in and whether the freedom of information leads to broad access to information \citep{Appelgren2018,Porlezza2019}. A third factor is the media market structure, the availability or lack of resources that allows experimenting with innovative formats of journalism that have not yet proven to be successful \citep{DeMaeyer2015}.

Data journalism is regarded as a form of content or genre innovation in journalism \citep{GarciaAviles2018,GarciaAviles2020}, which might provide news companies with an increased reputation or another competitive advantage. Data journalists use data sets and their own data analysis as sources for their stories, often bundled with infographics or interactive elements.

Media innovation nearly always contains some societal effect, as media reflects society in its content and organizational and technological structures \citep{Pavlik2000,Bruns2014}. When and how this innovation takes place is shaped by internal factors like staff incentives or leaders' behavior \citep{Paulussen2011,Ekdale2015,GarciaAviles2019}, but also by external influences, like technology changes, market opportunities or evolving industry norms, and audience behavior \citep{Anderson2013,Storsul2013,Bleyen2014,Ess2014}.

COVID-19 served as an accelerator for ongoing changes and innovation in journalism: the decline of print and other forms of ''traditional`` media, the rise of ''alternative`` news channels, restructured processes and altered skill requirements for journalists, changes in audience and their expectations and new approaches to journalism \citep{Quandt2022}. There is, however, some debate on the extent of these changes \citep{Hanusch2022}.

Most innovations during COVID-19 were developed in the product (like data visualizations or fact-checking), distribution (newsletters or podcasts), and commercialization (subscriptions and membership models). 

We will focus here on one innovation in particular: Science departments were more relevant and worked with data teams to create visualizations \citep{GarciaAviles2022}. Data journalists were central to these innovations, as they had the experience and technical means to create data visualizations \citep{Desai2021} and exploratory pieces on possible scenarios, which has increased awareness and accessibility to the numbers and fostered engagement of the audience \citep{Pentzold2021}. However, this also led to criticism about an ``information overload'' \citep{Krawczyk2021}, a ''bombardment`` with visualizations \citep{GarciaAviles2022} and a very small number of — often governmental — sources (\cite{Mellado2020,Aula2020}, see also \cite{Cawley2016,Tandoc2017}).

We will base this research on these empirical, qualitative observations and try to operationalize the relationship between data journalistic pieces — articles that were (co-)authored by a data journalist — and intra-newsroom cooperation between data and science journalists to be able to make quantitative conclusions about the extent that data journalism innovated during COVID-19. As a theoretical foundation, we will leverage the Communities of Practice model to help explain our findings.

\subsection{Communities of Practice}

At the beginning of the 1990ies, the focus in studies of social interaction started to move from individuals to groups. \citet{Zelizer1993} described journalism as interpretive communities ``united through shared discourse and collective interpretations of key public events'' \citep[p. 219]{Zelizer1993}. More prominently, \citet{Lave1991} developed the idea of Communities of Practice (CoP) as a social learning theory in organizations, ``a set of relations among persons, activity, and world, over time and in relation with other tangential and overlapping communities of practice'' \citep[p. 98] {Lave1991}. These communities exist in parallel to formal hierarchies of organizations as an informal social system. 

Central elements of these communities are a common domain of knowledge, a community caring about, and a shared practice to be effective in this domain \citep{Wenger2002}. The interactions within these communities are structured in three dimensions \citep{Wenger1998}: mutual engagement between community members fuelled by complementary or overlapping skills, mutual relationships that allow engagement. A joint enterprise is driven by negotiated responses to internal or external conditions, resources, or demands towards the community. And a shared repertoire in tools like language, routines, or tools that shape the work in the practice.

In later years, these basic sets have been amended by stages of community-building, which can be found in the field, but do not have to be passed through in this order \citep{Wenger2002}. This emphasizes the idea that CoP are dynamic and continually evolve through interactions and collaborations.

\begin{itemize}
\item Potential: when a first group of people starts to take on a certain topic. 
\item Coalescing: when the community starts to set up a basic structure.
\item Maturing: when the community grows and continues to increase and share knowledge.
\item Active: when there is an acceptable amount of members and the amount of added knowledge declines.
\item Dispersing: when the community is no longer relevant due to other sources or a loss of relevance for the domain.
\end{itemize}

The concept of Communities of practice has been applied throughout organizational research, also, on journalism. Journalists often form informal networks within and across newsrooms centered around shared beats, interests, or expertise. These communities influence professional identity, newsroom culture, and journalistic norms. The formation of CoP in journalism demonstrates the significance of collective learning, fostering a culture of continuous improvement and adaptation in the face of evolving media trends. \citet{Meltzer2017} characterized newsroom journalism as a distributed community of practice or many subcommunities that constantly cycles between phases as new technologies appear. We will use this concept to embed our findings into a theoretical framework that helps describe the collaboration of journalists from different editorial departments.

Data journalism, characterized by data-driven storytelling and visualization, has rapidly emerged as an essential practice in modern journalism. Within data journalism, CoP could potentially play a pivotal role in sharing technical expertise, discussing data analysis techniques, and disseminating innovative storytelling approaches. These communities empower data journalists to explore new storytelling methods, employ cutting-edge tools, and interpret complex datasets, ultimately elevating the quality and impact of data-driven reporting.

In response to the COVID-19 crisis, data journalism may have gained deeper inclusion in newsrooms, reflecting the increasing relevance of data-driven reporting in crisis communication. During this time, CoP could conceivably play a critical role in facilitating collaborations between data journalists and science journalists. By leveraging their shared expertise and resources, these communities may have contributed to producing accurate, visually engaging, and accessible information on the pandemic.

\subsection{Using Bylines to Study Journalism}

To operationalize these cooperations, we will use byline data.
Investigating bylines, the author attribution snippet above or below a journalistic article, to measure the implications of authorship have some tradition in journalism studies. They can mostly be distinguished by two aims: to show gender-related differences that the audience may perceive by looking at the author's name or to show the impact of computer-generated articles on the confidence and perception of the readers.

A famous initial study to measure gender-related attitudes with bylines is the work of Philip Goldberg, who showed in 1967 that female readers were likely to rate male authors more favorably than female authors \citep{Goldberg1967}. This work is often cited as a reference for bias against women authors. However, meta-research showed that the effect is negligible \citep{Burkhart1990,Swim1989}. Gender bias seems to be context dependent \citep{Dogruel2023}: Audiences still have a predefined vision of which areas females have to report on and rate them less credible if they leave those, for instance, sports journalism \citep{Klaas2022}, which at least indicates persistent marginalization of female bylines over 15 years \citep{Boczek2022}. However, \citet{Boczek2022} could not confirm readers' biases against female reporters.

Another use case for byline methods are studies about the perception of computer-generated text in journalistic articles. These are already used in various applications, like trading or sports reporting. Legacy newsrooms may use computer-generated articles as baseline reporting. They may enrich this by reporters' inputs and increase the value for users, which some journalists regard as a complement, not a replacement of their work \citep{Kunert2020}. Attributing authorship for automated content is difficult and raises ethical questions about the responsibility for news content and the requirements for transparency towards the readers \citep{Henrickson2018,Montal2017,VanderKaa2014,Graefe2016,Waddell2018}.

We will use bylines to measure the number of authors that contributed to an article and to allow for identification to which editorial department, mostly organized by subject area, a person belongs. Co-authorship analysis serves as a proxy for collaboration between journalists. However, it may not fully capture the intricacies of collaborative dynamics within newsrooms, as it does not account for informal exchanges and knowledge sharing that may occur without formal co-authorship.

\subsection{Research Questions}

The research questions outlined below will allow us to quantitatively explore the extent to which data journalism has proliferated in newsrooms during and after COVID-19, the changes in collaborative practices, and the role of science journalists in this context:\\

\textbf{Q1: Does the number of data journalistic pieces change after COVID-19?}
This research question examines whether there has been a measurable increase in the frequency of data journalistic pieces published in news outlets during and after the COVID-19 pandemic. By analyzing data journalism outputs over time, we seek to identify potential shifts in journalistic practices in response to the pandemic.\\

\textbf{Q2: Does the number of authors on data journalistic pieces change after COVID-19?}
This question explores whether there has been a change in the collaboration patterns among journalists working on data-driven pieces after the onset of the pandemic. Understanding how the number of authors involved in data journalism has evolved can provide insights into the intensification of collaborative efforts during times of crisis.\\

\textbf{Q3: Does the number of data journalistic pieces change across departments after COVID-19?}
In examining data journalism outputs across different departments within news organizations, we aim to assess whether the COVID-19 pandemic has influenced the distribution and shifted the focus of data-driven reporting.\\

\textbf{Q4: Does the authorship of science journalists change after COVID-19?}
This research question delves into the involvement of science journalists in data-driven reporting during and after the COVID-19 pandemic. By analyzing the authorship patterns of science journalists in data journalism, we aim to understand the role of scientific expertise in the intensified collaboration between data and science journalists in the context of the pandemic.\\

\textbf{Q5: Does prior common authorship change the probability of future co-authorships for data journalistic articles?}
Examining the co-authorship networks within data journalism, this question investigates whether prior common authorship experiences influence the likelihood of future collaborations. Understanding the dynamics of co-authorship relationships can provide valuable insights into collaboration and knowledge exchange patterns among data and science journalists during COVID-19.\\

By addressing these research questions, this study contributes to our understanding of the influence of COVID-19 on data journalism, shedding light on the changing landscape of journalism practice during times of crisis. The findings aim to advance scholarly knowledge on possibilities to generate quantitative insights in the field of computational communication science and journalism practice.

This article offers an initial quantitative analysis of the developments of data journalism in Germany during COVID-19 based on non-questionnaire data. We first present an overview of the literature on data journalism and the impact COVID-19 might have had on the practice. We then embed our findings within the existing research on communities of practice before presenting our results and discussing them in the light of the theory.

\section{Materials and Methods}

This research was conducted using computational methods to collect and analyze data. We will describe below how metadata on articles was collected on author pages and which statistical methods and network science models were used to derive the results.

\subsection{Data}

To identify data journalists, we started with a Slack group that was formed as an advocacy group for German data journalists by the non-governmental reporters' representation "Netzwerk Recherche" in the fall of 2020 \citep{NetzwerkRecherche2020}. While this may lead to potential self-selection bias, we assume the majority of data journalists to be members of this group, as there are no fees or further barriers, and participation in the group offers incentives, like discussions on current topics in the field, information on upcoming conferences or meet-ups, or a job market \citep{Witzenberger2022}. We further acknowledge that we are extracting data from a somewhat closed group to which one of the authors had access, which raises potential privacy concerns. However, as the main purpose of this study is to analyze the authorships of public media articles, this information is already publicly accessible elsewhere, but detecting data journalists would be a lot more challenging.

We limited data collection on the largest four journalism teams in Germany, Süddeutsche Zeitung (SZ), Spiegel, Zeit, and public broadcaster Bayerischer Rundfunk, and included the quite small data team of the Berlin-based newspaper Tagesspiegel to allow variance in team sizes, resulting in 688 articles from 363 distinct authors (7.99 percent data journalists, n = 29). 
This sample was limited due to the availability of historical article and author data and the effort of manual processes that needed to be performed to code and validate the data. We created a list of data journalists and aimed to collect all articles they were (co-)authoring by web-scraping the article metadata of the authors' pages provided by news media to showcase works by individual journalists. A data-journalistic article or piece is, therefore, an article that was (co-)authored by a data journalist. 
Data collection took place in November 2022 for articles between January 2019 and December 2021 to allow for somewhat similar periods pre- and post-COVID-19. The titles, authors' names, dates, and URLs for each article were collected.

Data preparation tool place in \citet{R2022} using the packages \citet{DplyR2023,Stringr2022,TidyR2023}. While title, dates, and URL information were easily parseable, the way of specifying authors' names differed greatly between organizations. In a few instances, Süddeutsche Zeitung only describes authorship by ``SZ-Autoren'' (''SZ-authors``), which did not yield any relevant information for our analysis on authors, but the article was used in the analysis of the prevalence of data journalistic articles. Newsroom departments were retrieved using the URL file path and clustered across media organizations.

The public broadcaster Bayerischer Rundfunk (BR) posed a further challenge to data access. As the German media landscape is split into private and public media, the organization of the latter is guarded by a legal contract between German federal states (``Interstate Treaty on Broadcasting and Telemedia'', \citet{RStV2019}). It restricts the time periods that editorial publications of public service broadcasting companies are available online. The authors retained a dataset of data journalistic articles created by the data team of Bavarian public broadcaster Bayerischer Rundfunk due to a personal request, which will be used in the analysis. 
However, we found no authorship-cooperation with science journalists for BR, limiting the data's meaningfulness for part of the research questions, looking for evidence to find increased cooperation, but still allowing for investigation of the prevalence of data journalistic articles.

As we only focus on data and science departments in this research, only authors from those two have been manually coded in the data, using self-descriptions on author pages, imprints, or descriptions of Twitter. We found that relations to those departments were very stable and distinct and did not change over time, which might be caused by the high specialization in science reporting or the work with data that is required.

\subsection{Methods}

To derive conclusions, this paper will be two-fold: We will use common statistical inferences and modeling tools, namely ${\chi}^2$ tests and negative binomial regression models, to validate changes in the number of articles and authors and deferrals between different departments. But to show the usefulness of analyzing journalistic cooperation with network analysis methods, we will then apply Relational hyperevent models (RHEM) to investigate which changes of authorship can be found between departments (to answer Q4 and Q5).

Relational hyperevent models are an advancement of relational event models (REM). A relational event is defined as a ``discrete event generated by a social actor and directed towards one or more targets'' \citep[p. 159]{Butts2008}. The central idea is to model the history of events to describe the probability rate for the next relational event, or to put it more bluntly: ``how and why do relational events happen?'' \citep[p. 183]{Pilny2016}. The data is collected in a longitudinal fashion and by design using time-stamped interactions, like e-mails or the publication data of articles in our case, which have been shown to give a more accurate reflection of interactions, compared with surveyal studies \citep{Corman1994}. This data can be enriched by adding individual attributes that may increase or hinder chances for interactions. 

REMs have been used for a variety of use cases, from the formation of friendships in a virtual social network \citep{Welles2014}, to study interactions between states \citep{Lerner2013}, or the network structure of successful Wikipedia article editing collaboration \citep{Lerner2019}. 

However, REMs require the network data to be in a dyadic format, taking the form of a source/sender and target/receiver relationship. Networks, like cooperation between multiple journalists to collectively write an article, require a different set of models, which can model interactions between one sender and multiple receivers or, as in our case, between multiple senders (the authors) and a single receiver (the article) — called `hyperedges' \citep{Kim2018}.

\citet{Lerner2019a} have proposed Relational hyperevent models, which can include the hyperedge structure in their output, and have shown its utility in studies on the network dynamics of contact diaries of former British Prime Minister Margarethe Thatcher \citep{Lerner2021}, and analysis of scientific coauthor networks \citep{Lerner2023}. 

\begin{figure}
\centering
\includegraphics[width=0.5\textwidth]{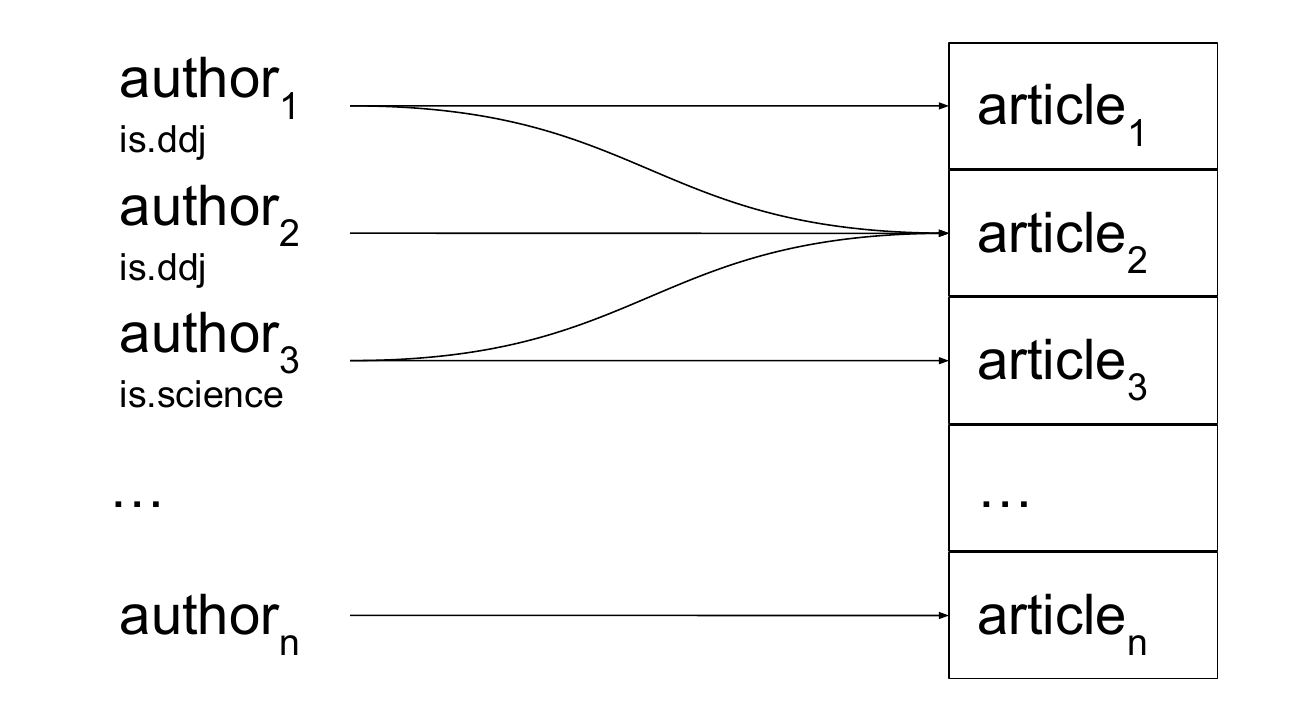}
\caption{Data model of network analysis showing an exemplary two-mode network between one or more authors connected to one article.} \label{fig:network_model}
\end{figure}

We will model our data as two-mode networks between a set of one or multiple authors and a set of single articles — shown in Figure \ref{fig:network_model} — with encoded information on the time of publication, former cooperations between authors, and their respective departments (data journalism, science, investigative or other). One author can only be connected once to a particular article, but multiple authors could be connected to an article, which would correspond to a co-authorship, resembling sources in the network. In our model, articles are modeled as targets of interactions, therefore, not connected to authors or other articles. A connection is, therefore, authorship between an author and an article. This allows us to investigate interactions between different authors over time and compare expected with actual article publications.

\section{Results}

\begin{figure}
\centering
\subfloat[Showing the published data journalistic articles per month, with periods of nationwide lockdowns highlighted.]{%
\resizebox*{0.45\textwidth}{!}{
\includegraphics{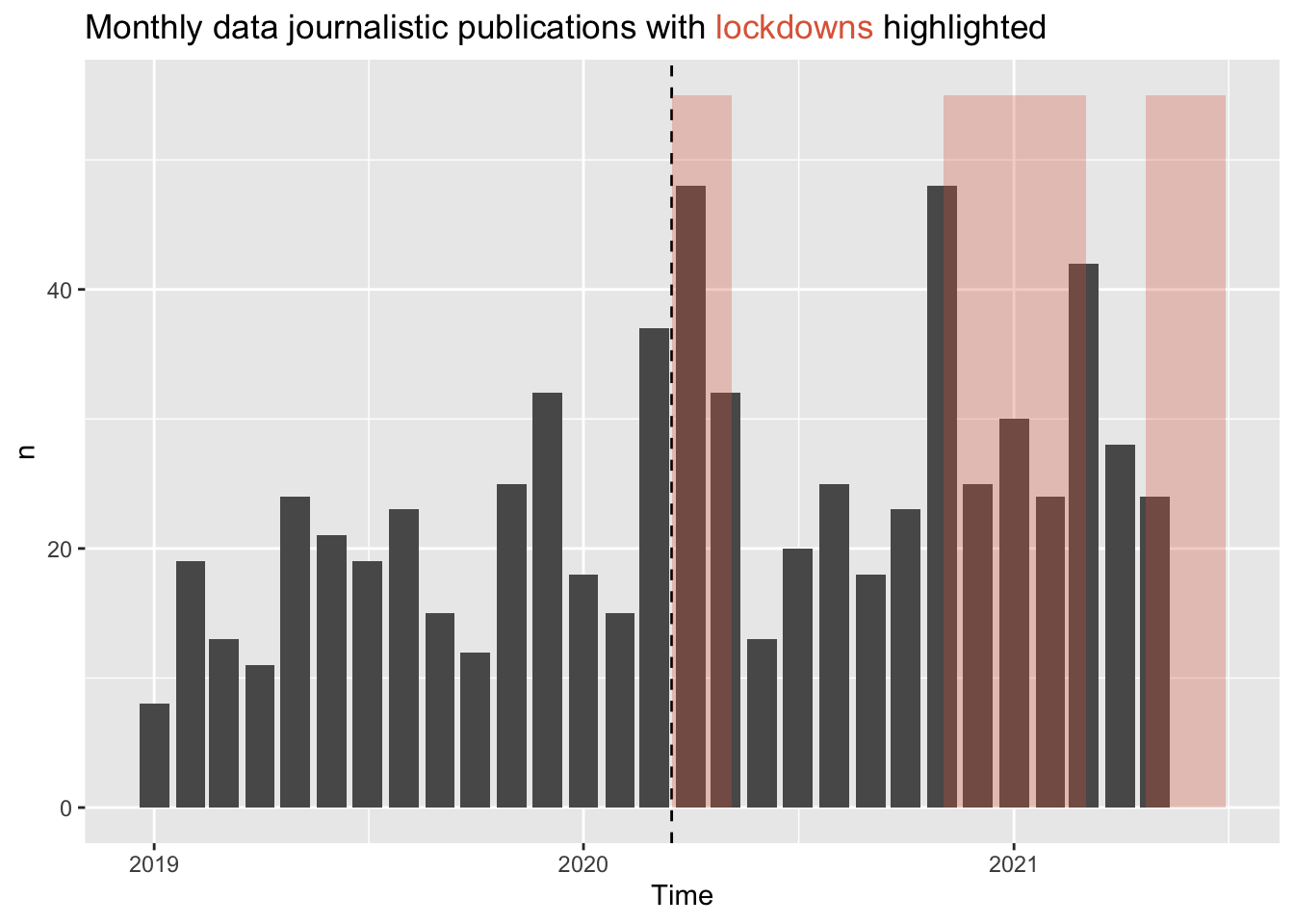}
\label{fig:monthly_publications_lockdowns}
}
}\hspace{5pt}
\subfloat[Showing the authorships per month, with periods of nationwide lockdowns highlighted.]{%
\resizebox*{0.45\textwidth}{!}{
\includegraphics{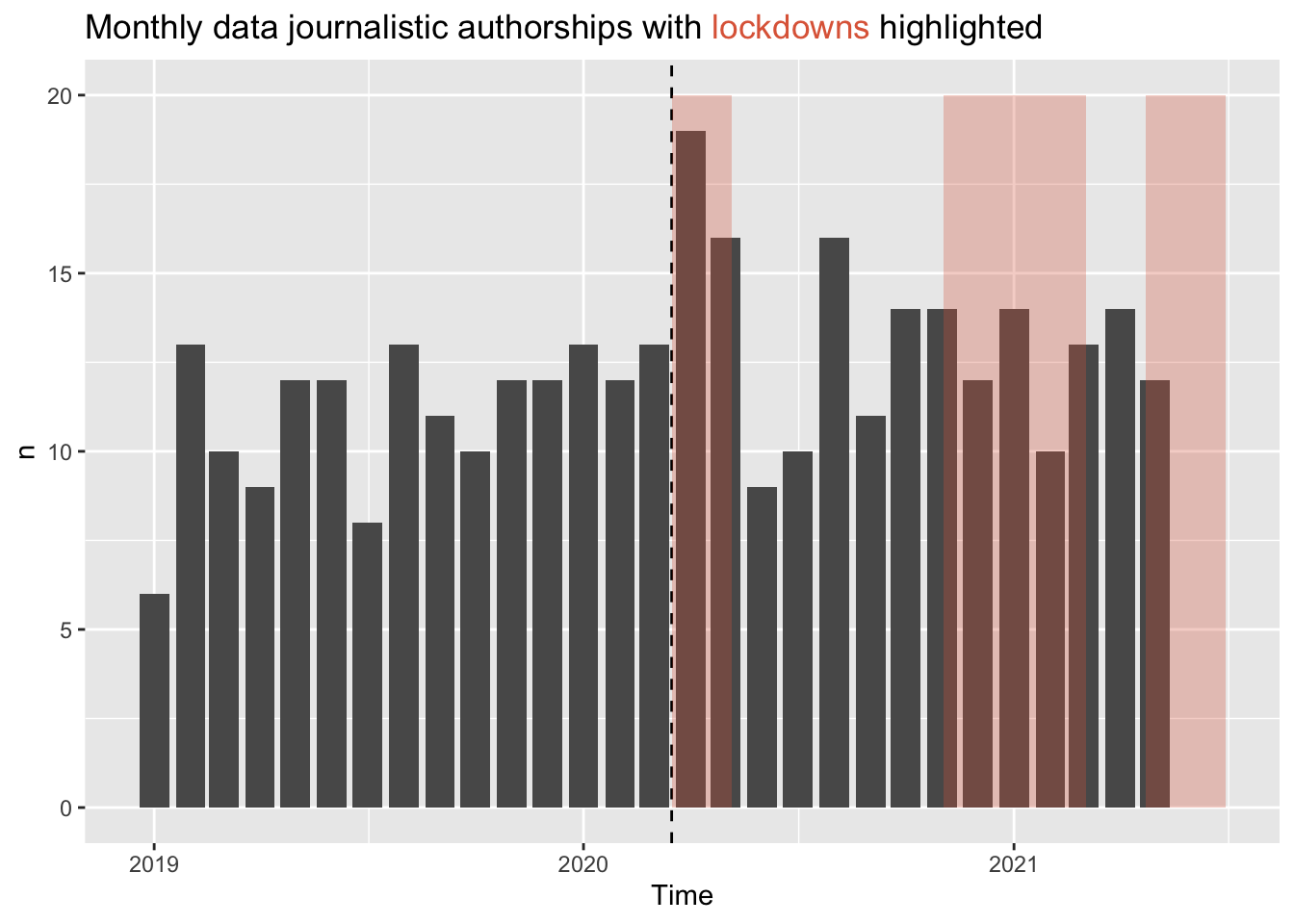}
\label{fig:monthly_authorships_lockdowns}
}
}
\caption{Comparing monthly number of publications (Fig. \ref{fig:monthly_publications_lockdowns}) and authorships (Fig. \ref{fig:monthly_authorships_lockdowns}).}
\end{figure}

We will now move on to lay out the results of the analysis. We retrieved 688 articles from 363 distinct authors between January 4th, 2019, to May 15th, 2021, for the five media companies. The time period was adapted to include periods before and past COVID-19 occurred, which we set to be March 16th, 2020, when the first lockdown was decided in Germany. Figure \ref{fig:monthly_publications_lockdowns} shows the monthly number of data journalistic articles with lockdowns highlighted, visualizing the increased publication rate in or close to lockdown periods.

\subsection{Some increase in articles, but not in authors}

\begin{figure}
\centering
\includegraphics[width=0.7\textwidth]{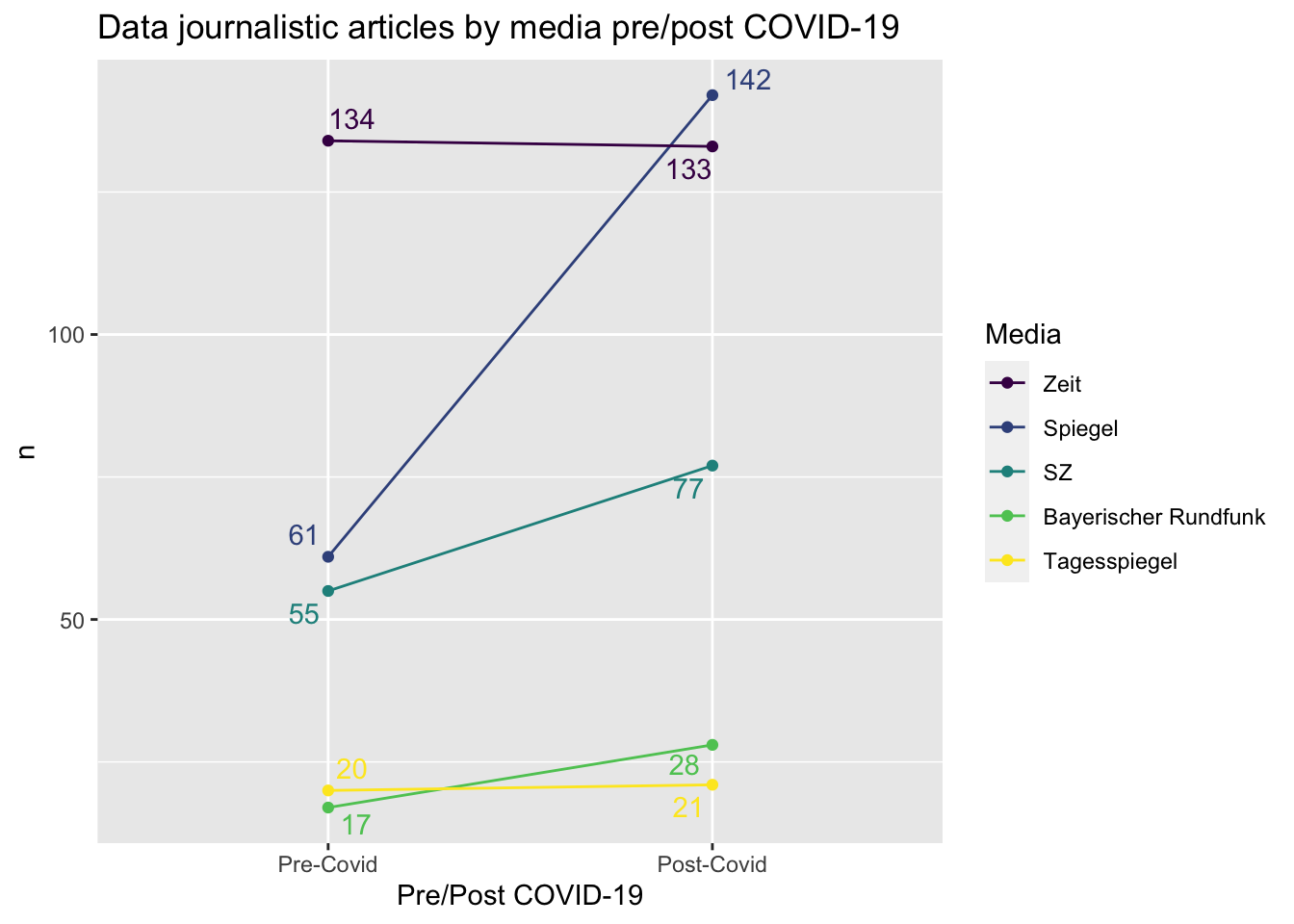}
\caption{Absolute counts of data-journalistic articles per media pre- and post-COVID-19.} \label{fig:articles_count_pre_post}
\end{figure}

Between the time before COVID-19 hit and after, we found a 40 percent increase of data journalistic articles across all observed media from 287 to 401 articles ($\chi^2 < 0.01$), which seems to affirm Q1. We observed these changes for three newsrooms: Spiegel, SZ, and Bayerischer Rundfunk. The counts for Tagesspiegel (20 vs. 21) remained nearly the same. For Zeit (134 vs. 133), they decreased slightly. Figure \ref{fig:articles_count_pre_post} displays these counts.

During the observation period, we found a large increase in publications during the initial lockdown in March 2020 across all media, and after a short decline, an evenly increasing number of publications, as shown in Fig. \ref{fig:monthly_publications_lockdowns}.

We then aimed to answer Q2, which shifted focus from publications to individual authors, to account for changes that might indicate an increased interest in the topic. Similar to the number of publications, we found an increase during the initial lockdown period but, in contrast, saw a decrease in the author numbers afterward when looking at Fig. \ref{fig:monthly_authorships_lockdowns}. This is backed up by results of a Wilcoxon signed rank test with continuity correction, which is used due to the sample's non-normal distribution, resulting in  $p>0.1$. A negative binomial regression, however, was used because the data indicated overdispersion, including media as a controlling factor, that yielded a 71 percent increase in author counts when holding all other variables constant (see Table \ref{tab:nb-authors}). While this indicates some increase in author numbers, it also points to editorial or team-specific differences between media.

\begin{table}
\tbl{Results of Negative Binomial regression model to predict the number of authors pre- and post-COVID, controlled by media.}
{
\begin{tabular}{@{\extracolsep{5pt}}lc} 
\\\toprule
 & \multicolumn{1}{c}{\textit{Dependent variable:}} \\ 
\\[-1.8ex] & n \\ 
\midrule \\[-1.8ex] 
  before\_covidPost-Covid & 0.535$^{***}$ \\ 
  & (0.159) \\ 
  & \\ 
 mediaSpiegel & 1.627$^{***}$ \\ 
  & (0.256) \\ 
  & \\ 
 mediaSZ & 1.242$^{***}$ \\ 
  & (0.258) \\ 
  & \\ 
 mediaTagesspiegel & $-$0.049 \\ 
  & (0.272) \\ 
  & \\ 
 mediaZeit & 2.474$^{***}$ \\ 
  & (0.253) \\ 
  & \\ 
 Constant & 3.635$^{***}$ \\ 
  & (0.210) \\ 
  & \\ 
\hline \\[-1.8ex] 
Observations & 10 \\ 
Log Likelihood & $-$50.471 \\ 
$\theta$ & 18.848$^{**}$  (9.180) \\ 
Akaike Inf. Crit. & 112.942 \\ 
\bottomrule\\[-1.8ex] 
\textit{Note:}  & \multicolumn{1}{r}{$^{*}$p$<$0.1; $^{**}$p$<$0.05; $^{***}$p$<$0.01} \\ 
\end{tabular}
}
\label{tab:nb-authors}
\end{table}
 
\subsection{Science department becomes data journalistic}

We then set to compare the prevalence of data journalistic articles across different departments in the newsroom. Departments were retrieved from the URL subdirectories of the articles and manually bucketed (i.e., politics, business, science, arts/culture) to align different naming conventions where a general link could be made. We investigated this question in two ways: First, we ran a negative binomial regression to investigate which department data-journalistic articles were presented pre- and post-COVID-19. The model (Table \ref{tab:nb-departments}) found negative effects on data journalistic publications, digital and opinion departments, but a very strong positive effect on science with a 175 percent increase, answering Q3 and Q4.

\begin{table}
\tbl{Results of Negative Binomial regression model to predict the number of publications pre- and post-COVID, controlled by department.}
{
\begin{tabular}{@{\extracolsep{5pt}}lc} 
\\\toprule
 & \multicolumn{1}{c}{\textit{Dependent variable:}} \\ 
\\[-1.8ex] & n \\ 
\midrule \\[-1.8ex] 
  before\_covidPost-Covid & 0.793$^{***}$ \\ 
  & (0.074) \\ 
  & \\ 
 departmentbusiness & $-$0.761$^{***}$ \\ 
  & (0.185) \\ 
  & \\ 
 departmentdigital & $-$1.001$^{***}$ \\ 
  & (0.216) \\ 
  & \\ 
 departmentlocal & $-$0.569$^{***}$ \\ 
  & (0.199) \\ 
  & \\ 
 departmentmobility & $-$0.757$^{***}$ \\ 
  & (0.213) \\ 
  & \\ 
 departmentopinion & $-$1.580$^{***}$ \\ 
  & (0.369) \\ 
  & \\ 
 departmentother & $-$0.427$^{**}$ \\ 
  & (0.210) \\ 
  & \\ 
 departmentpolitics & 0.702$^{***}$ \\ 
  & (0.123) \\ 
  & \\ 
 departmentscience & 1.169$^{***}$ \\ 
  & (0.117) \\ 
  & \\ 
 departmentsports & $-$0.651$^{***}$ \\ 
  & (0.197) \\ 
  & \\ 
 departmentwork & $-$0.699$^{**}$ \\ 
  & (0.279) \\ 
  & \\ 
 Constant & 1.969$^{***}$ \\ 
  & (0.116) \\ 
  & \\ 
\hline \\[-1.8ex] 
Observations & 68 \\ 
Log Likelihood & $-$421.564 \\ 
Akaike Inf. Crit. & 867.127 \\ 
\bottomrule\\[-1.8ex] 
\textit{Note:}  & \multicolumn{1}{r}{$^{*}$p$<$0.1; $^{**}$p$<$0.05; $^{***}$p$<$0.01} \\ 
\end{tabular}
}
\label{tab:nb-departments}
\end{table}

To further investigate and examine the predictive capabilities of Relational hyperevent models, we included data, science, and investigative departments as attributes into an RHEM to investigate the probability of cooperation happening between journalists from these departments.

The three departments, data journalism, science journalism, and investigative journalism, were specifically modeled, as these were the ones that authors had been coded to before. All journalists in the data were regarded as available co-authors for the model. We used a so-called conditional size-directed hyperedge observation for the model, which samples events with non-events on a given time frame. While this does not allow us to model pre- and post-COVID-19 prevalences, it allows us to observe the authors' departments. These models were created for each media company and analyzed using the Cox proportional hazards model \citep{Cox1972} that originates in survival modeling and is used to relate time, occurrences of events, and further co-variables. As Table \ref{tab:eventnet-coxph-departments} shows, we find some evidence for increased participation in data journalistic articles by science journalists. This is measured by the interaction $sender.avg.science:post\_cov$, which measures the occurrence of science journalists publishing with data journalists after COVID-19. This effect is not visible for data journalists, who seem to co-publish together very often as variable $sender.avg.ddj$ indicates.

\setcounter{table}{2}
\begin{sidewaystable}
\tbl{Results of Relational hyperevent model.}{
\begin{tabular}{@{}lccccc@{}}
\toprule
 & \multicolumn{5}{c}{\textit{Dependent variable:}} \\ 
\cline{2-6} 
\\[-1.8ex] & \multicolumn{5}{c}{Article Publication} \\ 
 & \multicolumn{1}{c}{SZ} & \multicolumn{1}{c}{SPIEGEL} & \multicolumn{1}{c}{BR} & \multicolumn{1}{c}{ZEIT} & \multicolumn{1}{c}{TAGESSPIEGEL} \\ 
\\[-1.8ex] & \multicolumn{1}{c}{(1)} & \multicolumn{1}{c}{(2)} & \multicolumn{1}{c}{(3)} & \multicolumn{1}{c}{(4)} & \multicolumn{1}{c}{(5)}\\ 
\midrule
 sender.avg.ddj & 4.970$^{***}$ & 9.250$^{***}$ & 4.035 & 3.308$^{***}$ & 4.917$^{***}$ \\ 
  & (0.960) & (1.185) & (3.021) & (0.459) & (1.093) \\ 
  & & & & & \\ 
 sender.avg.science & -0.877 & -0.590 &  & 0.171 & -2.315 \\ 
  & (1.707) & (1.867) & (0.000) & (0.874) & (2.525) \\ 
  & & & & & \\ 
 sender.avg.investigative & 1.448 & -3.764 & 4.870 & 1.427$^{**}$ &  \\ 
  & (2.548) & (3.210) & (3.022) & (0.682) & (0.000) \\ 
  & & & & & \\ 
 sender.sub.rep.1 & 6.333$^{***}$ & 2.474$^{***}$ & 1.790 & 2.974$^{***}$ & 2.274 \\ 
  & (1.615) & (0.924) & (4.171) & (0.305) & (2.719) \\ 
  & & & & & \\ 
 sender.sub.rep.2 & -2.826 & 1.864 & 2.888 & 3.161$^{***}$ & -3.161 \\ 
  & (1.754) & (3.012) & (4.065) & (0.964) & (18.278) \\ 
  & & & & & \\ 
 closure & 0.042 & -1.082$^{**}$ & 0.280 & 0.016$^{***}$ & -0.520 \\ 
  & (0.071) & (0.492) & (0.219) & (0.006) & (0.733) \\ 
  & & & & & \\ 
 closure:post\_cov & -0.044 & 1.069$^{**}$ & -0.263 & -0.009 & 0.474 \\ 
  & (0.071) & (0.492) & (0.223) & (0.006) & (0.744) \\ 
  & & & & & \\ 
 sender.sub.rep.1:post\_cov & -3.787$^{**}$ & -0.591 & 0.622 & -0.713$^{*}$ & 4.634 \\ 
  & (1.718) & (0.999) & (4.886) & (0.418) & (4.942) \\ 
  & & & & & \\ 
 sender.sub.rep.2:post\_cov & 3.628$^{*}$ & 1.490 & 0.812 & 0.110 & 1.278 \\ 
  & (1.945) & (3.166) & (4.711) & (1.371) & (18.829) \\ 
  & & & & & \\ 
 sender.avg.ddj:post\_cov & 1.112 & -2.521$^{*}$ & -0.248 & -0.038 & 0.357 \\ 
  & (1.272) & (1.357) & (3.720) & (0.642) & (1.930) \\ 
  & & & & & \\ 
 sender.avg.science:post\_cov & 4.309$^{**}$ & 3.994$^{**}$ &  & 2.195$^{**}$ & 2.504 \\ 
  & (1.930) & (1.992) & (0.000) & (1.048) & (3.204) \\ 
  & & & & & \\ 
 sender.avg.investigative:post\_cov & -0.931 & 0.197 & -2.184 & -0.260 &  \\ 
  & (3.987) & (4.041) & (3.749) & (1.011) & (0.000) \\ 
  & & & & & \\ 
\hline \\[-1.8ex] 
Observations & \multicolumn{1}{c}{9,393} & \multicolumn{1}{c}{12,928} & \multicolumn{1}{c}{2,929} & \multicolumn{1}{c}{22,725} & \multicolumn{1}{c}{2,121} \\ 
R$^{2}$ & \multicolumn{1}{c}{0.050} & \multicolumn{1}{c}{0.051} & \multicolumn{1}{c}{0.010} & \multicolumn{1}{c}{0.040} & \multicolumn{1}{c}{0.026} \\ 
Max. Possible R$^{2}$ & \multicolumn{1}{c}{0.088} & \multicolumn{1}{c}{0.090} & \multicolumn{1}{c}{0.087} & \multicolumn{1}{c}{0.090} & \multicolumn{1}{c}{0.087} \\ 
Log Likelihood & \multicolumn{1}{c}{-191.765} & \multicolumn{1}{c}{-266.247} & \multicolumn{1}{c}{-119.031} & \multicolumn{1}{c}{-613.324} & \multicolumn{1}{c}{-69.244} \\ 
Wald Test & \multicolumn{1}{c}{221.830$^{***}$ (df = 12)} & \multicolumn{1}{c}{289.290$^{***}$ (df = 12)} & \multicolumn{1}{c}{28.590$^{***}$ (df = 10)} & \multicolumn{1}{c}{609.420$^{***}$ (df = 12)} & \multicolumn{1}{c}{44.190$^{***}$ (df = 10)} \\ 
LR Test & \multicolumn{1}{c}{480.407$^{***}$ (df = 12)} & \multicolumn{1}{c}{682.140$^{***}$ (df = 12)} & \multicolumn{1}{c}{29.615$^{***}$ (df = 10)} & \multicolumn{1}{c}{925.125$^{***}$ (df = 12)} & \multicolumn{1}{c}{55.347$^{***}$ (df = 10)} \\ 
Score (Logrank) Test & \multicolumn{1}{c}{1,772.893$^{***}$ (df = 12)} & \multicolumn{1}{c}{2,144.024$^{***}$ (df = 12)} & \multicolumn{1}{c}{88.678$^{***}$ (df = 10)} & \multicolumn{1}{c}{4,624.465$^{***}$ (df = 12)} & \multicolumn{1}{c}{81.232$^{***}$ (df = 10)} \\ 
\bottomrule
\textit{Note:}  & \multicolumn{5}{r}{$^{*}$p$<$0.1; $^{**}$p$<$0.05; $^{***}$p$<$0.01}
\end{tabular}}\label{tab:eventnet-coxph-departments}
\end{sidewaystable}

\subsection{Authorship collaborations change}

An additional advantage of modeling the authorship events as a probabilistic network is the ability to gain deeper insights into the cooperation between authors. Using the RHEM model (see Table \ref{tab:eventnet-coxph-departments}), we could analyze subset repetitions. Subset-repetition (modeled as sender.sub.rep) describes the probability of exact or partly identical authors across several articles. In our case, we limited the measure to subsets of size one — meaning two authors — and two — a subset of three authors, as we did not see any larger subsets. 

The RHEM model showed that there was generally a significant subset repetition of size one for previous co-authorships between authors in SZ, Spiegel, and Zeit, indicating that authors tend to work together in smaller, prior constellations, affirming Q5.
For Zeit, this also included subsets of three authors, indicating repeated cooperation between multiple journalists. For SZ, we only found an increase for those sets of three authors in the interaction with the post-COVID-19 variable.

However, for some outlets, we also found negative effects on the $sender.sub.rep.1:post\_cov$ interaction. This could indicate changes in author bylines after COVID-19 as the probabilities for partial subset repetitions of prior cooperation decreased. In short, existing cooperations might have been discontinued due to the changes in interests that COVID-19 brought.

A second variable to investigate authors' cooperations, called $closure$, was only small and significant for ZEIT and negative for Spiegel. It indicated the probability of co-authoring with another journalist if that journalist has already published together with a common third author. This triadic closure is common in social networks \citep{Granovetter1973}. Interestingly, while it was negative for Spiegel overall, the interaction with the post-COVID-19 time period turned this into an increase, opening an interpretation that this effect was initiated during the pandemic situation.

\section{Discussion}

The reporting on COVID-19 was data-driven. News media published visualizations on the prevalence and effects of the COVID-19 pandemic worldwide. Therefore, it is only a small step to argue that data journalism played a crucial role in enabling newsrooms to prepare and provide these charts and dashboards \citep{Pentzold2021,Quandt2022}. However, the extent to which this has led to new or increased cooperation between data journalism teams and other areas of the newsroom has not been studied thoroughly \citep{GarciaAviles2022}.

As we have laid out in the Literature Review section, it might be worthwhile to ground this research into the existing Communities of Practice (CoP) literature to be able to regard the results in light of the theory of learning in organizations. The cooperation between data journalists and other departments can be viewed in this perspective as a Community of Practice between data specialists and science experts. As we have seen, there was a rapidly increasing number of data journalistic publications right at the start of the pandemic (Q1). This can be described as some maturing phase of communities \citep{Wenger2002}, which led into an active phase thereafter. The surge in data journalism during this crisis reflects the dynamic engagement within CoP, as data journalists leveraged their shared expertise to respond to the demand for data-driven reporting. As we have not seen a decrease in our data — there is some indication that the level of investment into the cooperations has not declined. The persistently high number of publications indicates that there is a general audience interest in data journalistic work or a generally higher acceptance of the work within the newsroom. Both are classical examples of exogenous factors that drive and shape a joint enterprise in a Community of Practice.

Admittedly, we see a decline in authorships right after the first peak in the first lockdown (Q2), which seems to contradict this argument. However, there has to be a distinction between the authors and the publications. COVID-19 has drawn great interest from news audiences, which might have caused journalists from different departments to cooperate with data journalists initially. After the dust began to settle, those might have returned to their respective departments and abandoned the data journalistic partnerships. In Community of Practice phases, this can be described as a piece of evidence for the dispersing phase of communities, when members leave the community due to loss of relevance. However, as \citet{Meltzer2017} pointed out, journalists tend to be members of different Communities of Practice. In our case, cooperation between data and science journalists has increased greatly.

Further research may investigate longitudinal effects on the prevalence of data-driven journalism and whether this explicitly takes place in cooperation with data journalism departments or has become detached and is now implicitly included in subject-specific editorial areas.

With the help of Relational hyperevent models (RHEM), we increased our understanding of journalistic author networks. In general, we found a strong sender subset repetition for data journalists, indicating that data journalists tend to cooperate with other data journalists. This aligns with the Communities of Practice understanding of creating a joint enterprise for a common domain of knowledge, a manifestation of continuous knowledge exchange and collaborative practices.

Interestingly, probabilities for sender subset repetition were negatively influenced by a post-COVID-19 dummy variable, which points to the possibility that COVID-19 has reshaped co-authorships in data journalism for some media companies. This observation is reinforced by a highly increased probability of publications featuring a science journalist after COVID-19, which points to a change through the pandemic. A new community was formed caring about the explainability of the pandemic activity through the use of scientific data and visualizations. This adaptation demonstrates the flexibility and adaptive capacity of CoP in journalism.

Significant effects on network closures were observed for ZEIT and Spiegel — and for those in two different directions. Network closures describe the probability that prior co-authorship of A and B with a third author, C, might increase chances for A and B to also cooperate in the future. One could argue that prior common authorship between a data journalist and a science journalist might also lead to a common publication with another data journalist and the same science journalist. While this might be true for ZEIT, the case for Spiegel is twofold. The number is below zero, indicating a lack of closure tendencies. However, after COVID-19, this probability is positive, which points to an increased chance for new cooperation. In combination with another observation from Spiegel on the decrease of data journalistic co-authorships after COVID-19, this could indicate less intra-data journalistic cooperation but increased co-authorships with other departments, particularly science (Q5).

The cooperation between data and science journalists can be described theoretically as combining different `transactional expertise' from data to science. Journalists have been shown to be reluctant to communicate and convey uncertainty when reporting on scientific results \citep{Witsen2019}. When observations of official measurements contradict public experience, this may lead to an alienation between people and statistics, whereas some journalists may rely heavily on quantification and may blindly trust the numbers \citep{LugoOcando2017}. This led to calls for more direct reporting on the uncertainty in reporting \citep{Anderson2018} or to enable journalists with `transactional expertise,' which defines knowledge that enables an individual to converse about a certain topic without being able to practically work within the field \citep{Collins2004,Witsen2018}. The complementary combination of data and visualization skills and the ability to understand and report on scientific research can be combined together in a Community of Practice to fuel the journalistic output during the pandemic reporting, leading to increased publications of science and data journalists (Q3 and Q4).

To summarize, we found the number of data-journalistic articles changed between -0.7 and 133 percent (overall:  percent) for similar time periods before and after COVID-19 hit, with four of five researched media increasing their numbers, and all media having high values for the time of the initial two lockdowns in March 2020, and Winter 2021. We also saw a huge decrease in publications across nearly all newsroom departments, with science being the sole outlier that increased its share clearly. This effect was also observable using RHEMs, which, in addition, indicated changes in the co-authorship structure pre- and post-COVID-19. At the same time, we found, generally, indications of subset repetition, which suggests recurring publication with previously co-authoring journalists. The appearance of COVID-19 reversed this effect, which implies the creation of new authorships.

\subsection{Limitations}

A number of factors limit this research. It is focused on a subset of the German data journalistic media landscape, covering the largest players in data journalism but leaving out most public broadcasters that do not provide author pages that could be scraped in a similar manner as private-owned media companies do. The study's analysis is based on data from five German newsrooms, which may limit the generalizability of the findings to a broader journalistic context. While the sample provides valuable insights into data journalism practices during COVID-19, caution should be exercised when extrapolating the results to other regions or news organizations with different characteristics and practices. The focus on Germany is also narrowing the view on a Western democracy, where data journalism has already well over a decade of history, and access to data is easier to achieve than it might in authoritarian political regimes, where COVID-19 might have also played a different role in public discourse.

A further limitation of the data is the perspective from which it was taken. During our observation period, we used the author pages of data journalists to build a ground dataset of data journalistic articles. This, however, defines data journalism as the work of data journalists, which were taken from a Slack channel, as described in the methods section. This implies that data journalistic work done completely by non-data journalists would not be included in our dataset. We expect this to be a very small number due to the focus on specific skills that might be bundled together in specific teams, which in turn form networks with other data journalists and should be visible in channels like the Slack group. In order to streamline our research efforts and allocate resources efficiently, we designated the coding of data to specifically target departments focused on journalism, science, and investigative reporting. We also observed far fewer department changes, as we suspected would happen in other areas of the newsroom and would affect the modeling effort.

Another effect on the prevalence of data journalism in Germany might have been the federal elections in 2021, as elections have traditionally been an important season for data journalists, which might have kept the number of publications higher than they might have been towards the end of the COVID-19 pandemic. However, we did not find a large decline in science data journalism towards the end of our observation period, just a small increase in political data journalism, which might indicate the expressed influence but also shows the limited extent it had. The research focuses on data journalism during the initial phase of the COVID-19 pandemic. It is essential to recognize that the pandemic's impact on data journalism and collaboration dynamics may continue to evolve over time. Future research could consider conducting longitudinal studies to examine how these trends develop over extended periods.

\section{Conclusion}

COVID-19 has influenced industries around the world, such as journalism. Data journalism especially came to increased attention, as many parts of pandemic reporting were based on data and visualizations for which data journalists had tools and knowledge. However, they also gained standing inside the newsroom and increased cooperation with science departments. We have analyzed co-authorships of German data journalists across five newsrooms.

We found that there was a significant increase of data journalistic pieces for most researched media during COVID-19, leading to more articles published, especially in scientific departments; the average number of authors per article also slightly increased during the initial phase of the pandemic, but since then decreased slightly. We found evidence of general recurring cooperation between previous (data journalistic) co-authors, which the occurrence of COVID-19 negatively influenced, which led to new, increased cooperation between data and science journalists and an increased number of publications in science departments during the pandemic.

The findings suggest that Communities of Practice play a vital role in facilitating collaborations, knowledge exchange, and innovation, enabling newsrooms to adapt to rapidly changing circumstances and produce credible data-driven reporting during challenging times. As journalism continues to evolve, the dynamics of Communities of Practice offer valuable insights for news organizations seeking to enhance journalistic cooperation.

\section*{Acknowledgements}

The authors would like to thank Termeh Shafie and Juergen Lerner for their valuable input and patient feedback on the network models and Florian Stalph for his feedback on the theoretical embedding of the work.

\section*{Declaration Of Interest Statement}

There are no potential conflicts of interest to disclose.

\section*{Data availability statement}

The data that support the findings of this study are openly available in "Authorships German Data Journalists 2019-2021" at https://doi.org/10.7910/DVN/AGTEVS.	


\bibliographystyle{apacite}
\bibliography{expose}

\end{document}